\documentclass[aps,showpacs,prstper,reprint,groupedaddress]{revtex4-1}
\usepackage{graphicx,dcolumn}

\begin{document}


\title{Diversity in Physics Education Research:  A Decade of Inclusion and Future Challenges}


\author{Laszlo Frazer}
\email[]{per@laszlofrazer.com}
\affiliation{School of Chemistry, UNSW Sydney, NSW 2052 Australia}
\affiliation{Department of Chemistry, Temple University, 1901 N. 13th St.  Philadelphia, PA 19122 USA}

\date{\today}

\begin{abstract}
Diversity has an important, multifaceted role in the success of education.  I have used a corpus analysis of the first ten years of \textit{Physical Review Physics Education Research} to investigate the diversity, broadly defined, of physics education research.  The results show that the physics education research community has extensively investigated the diversity of physics students, especially with respect to gender.  However, some less visible but still numerous groups of students are little studied.  On the whole, the contents of \textit{Physical Review Physics Education Research} reveals a robust research community in physics education, but there is room for growth with respect to working with different kinds of students and the interactions between student attributes and teaching methods.
\end{abstract}


\maketitle

\section{Introduction}
Diversity is important to the effectiveness and equity of education.  To identify diversity-related successes and future challenges, this article seeks to semiquantitatively meta-analyze the role of diversity in physics education research.  Using an artificial intelligence technique and brute-force searches, I show that some ideas (such as concepts or gender) are very popular in physics education research while others (such as multiple intelligence) are not. In addition to measuring research on student diversity, I measure methodological choices made by physics education researchers within the context of the effect that methodological choices have on ways diversity can be studied.

Previous metaresearch has investigated how people come to be physics education researchers \cite{van2014educational} and faculty familiarity with results from physics education research \cite{henderson2009impact}.   A report on the origin and evolution of physics education research was created for the United States National Academies \cite{cummings2011developmental}.  \textit{Physical Review} editors have made some statistics available about their other journals \cite{prchina},  while editors of physics education research papers have reported on their editorial practices \cite{hsu2007publishing}.

In this article, I used two approaches to meta-analyze the papers published in \textit{Physical Review Physics Education Research}.  The first approach uses an artificial intelligence algorithm to determine the main ideas of physics education research.  This approach does not require the researcher to make any choices of subject matter.  The second approach searches the text directly, a method that is more sensitive to the presence of minor ideas but only investigates topics that have been explicitly selected.
\section{Methods}
\textit{Physical Review Physics Education Research} is a ``peer-reviewed online open-access journal'' first published in 2005.  
It is easy to use as a corpus of text because it has highly consistent copy editing.  
Compared to some other journals, it tends to have a greater focus on the fundamentals of physics education research, with little material written for education practitioners or students as opposed to education researchers.  
Some related resources that were not analyzed include \textit{The Physics Teacher}, \textit{American Journal of Physics}, \textit{Physics Education Research Conference Proceedings}, and \textit{arXiv physics.ed-ph}.
\subsection{Suffix Tree Clustering Analysis}

Suffix Tree Clustering is an algorithm \cite{osinski2005carrot2} for automatic document classification.  From the perspective of meta-analysis of scientific literature, its advantages over other clustering methods include the ability to identify phrase cluster labels, the ability to merge similar document clusters, and good computational performance.  The algorithm also requires no training in the content of the documents that are classified, giving the algorithm an element of objectivity.  However, it does use lexical information about the document language and subjectively adjustable scoring parameters.

\begin{figure*}
 \includegraphics[width=\textwidth]{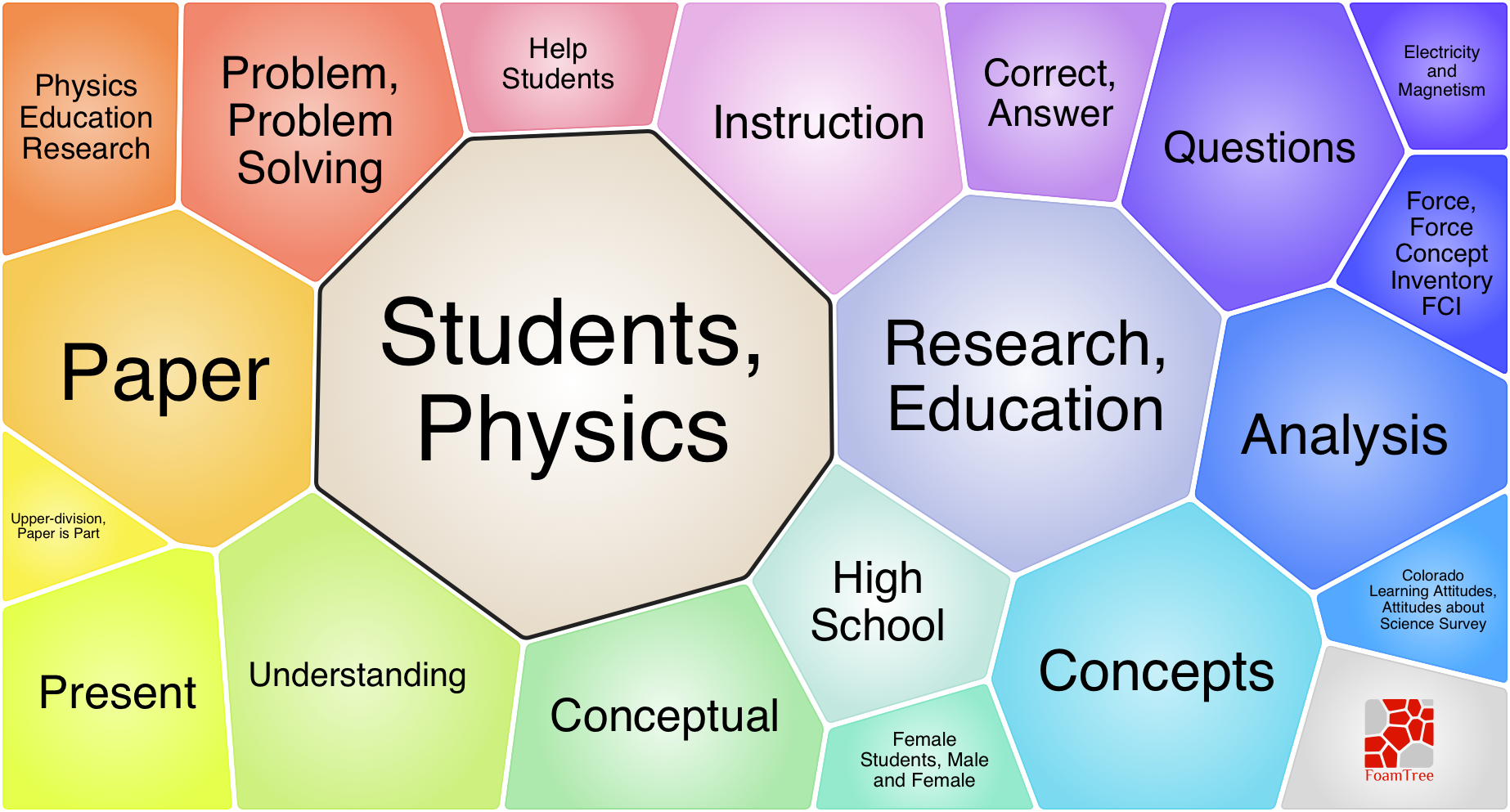}
 \caption{\label{fig:clusters} The 20 top-scoring clusters of abstracts in \textit{Physical Review Physics Education Research}, presented as a ``Foamtree'' Voronoi diagram.  The cluster area is determined by the number of papers in each cluster.  Articles and labels may be assigned to multiple clusters.}
 \end{figure*}

Document clusters in \textit{Physical Review Physics Education Research} are displayed in Fig. \ref{fig:clusters}.  For performance reasons, clustering was undertaken using only the abstracts (n=342)  of the first eleven volumes of the journal.  The Carrot$^2$ implementation of the algorithm was used \cite{osinski2005carrot2,stefanowski2003carrot2}.  Only the 20 highest scoring clusters were considered in this analysis.  These range in size from ``Female Students, Male and Female'' (26 abstracts) to ``Students, Physics'' (341 abstracts).  The threshold of 20 clusters was set subjectively because I could not identify a valid objective threshold.  Documents can appear in multiple clusters, labels can appear on multiple clusters, and not all labels need appear in every document in a cluster.

Document clustering provides an excellent overview of widespread concepts in physics education research.  Since it operates relatively objectively, it has limited utility in meta-analysis, since it may not identify any clusters that are relevant to a particular goal.  In a sense, the greatest insights can come from identifying concepts that are missing from the top cluster labels; these are ideas that do not have a strong research community or are not deemed important enough to appear in abstracts.  

\subsection{Terminology Frequency Analysis}
To investigate the presence of concepts not found in the top 20 document clusters, the full text  of \textit{Physical Review Physics Education Research} was converted to ASCII format and searched on a case-insensitive basis.  Errata, figures, and supplementary information were excluded but references, short papers, editorials, and announcements were included.  Since the titles of references are included in the reference list, a key assumption here is that ideas mentioned in the titles of references are relevant to the citing paper.

\begin{figure}
 \includegraphics[width=.5\textwidth]{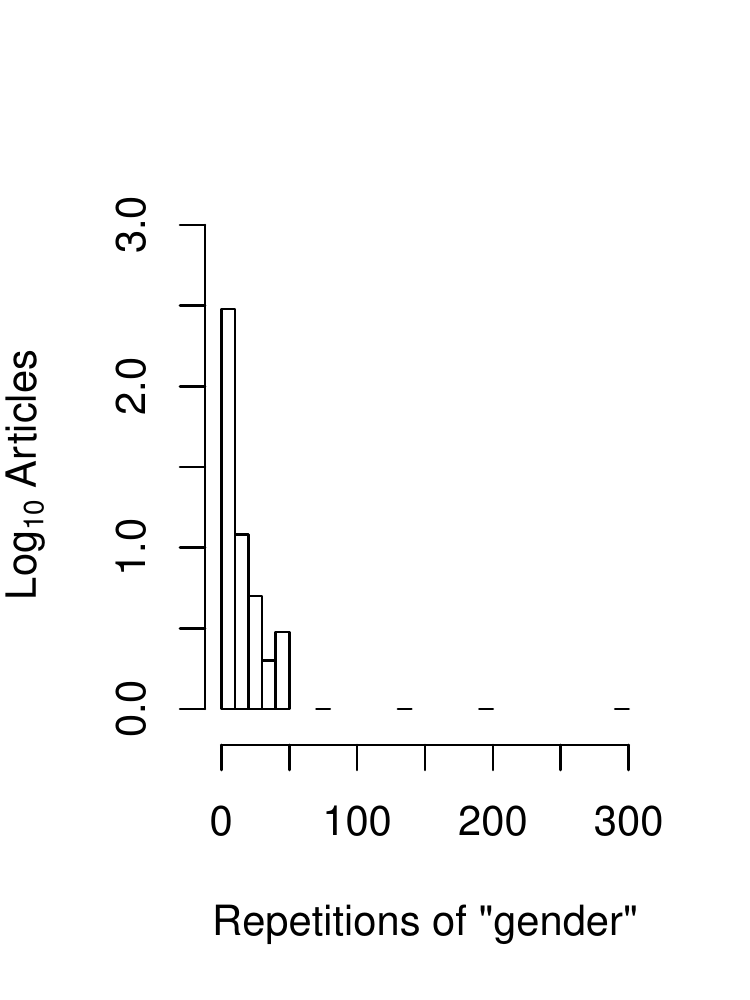}%
 \caption{\label{fig:hist} A logarithmic histogram showing the skewness of the number of occurrences of \textit{gender} in articles.}
 \end{figure}

Search string frequency can be highly skewed.   Figure \ref{fig:hist} shows an example for \textit{gender}.  Ref. \onlinecite{madsen2013gender}, a review article, mentions \textit{gender} 294 times, for a word frequency 500 times higher than the frequency of \textit{gender} in the Google Ngrams English corpus.  \textit{Gender} is not included in 243 articles.  Based on this skewness, I chose to count articles containing a given string, treating articles as the basic unit of research.  Articles typically contain $10^4$ words.
 
Terminology frequency analysis has several limitations.  It is up to the researcher to choose the strings of text to investigate.  In addition, natural language is subject to ambiguities such as synonymy and homography.  In the case of synonymy, it is necessary to try to identify multiple related words, such as ``remember,'' ``memorize,'' and ``recall.''  Homographs that could not be disambiguated were not analyzed.  For example, ``class'' is particularly difficult because it occurs in ``classroom,'' ``classical mechanics,'' ``Colorado Learning Attitudes about Science (CLASS),'' and ``socioeconomic class.''  Some strings unexpectedly appear in the names of authors.

In this article, \textbf{bold} text represents an inclusive OR search for strings in the categories listed in Appendix \ref{wl}.  For example, if an article contains both ``doctoral'' and ``postgraduate,'' it would count once towards the \mbox{\textbf{postgrad}} category.  \textit{Italic} text is a search string.  Text in \{braces\} is example context that shows what a string was intended to search for.  As an example, a search for \textit{analy}\{sis\} would count ``analysis,'' ``analyze,'' and ``reanalyze.''

\section{Student Populations\label{sec:pop}}

Student populations research encompasses the variety and commonality of students' personal traits.  Student diversity is highly valued for its role in the equity and efficacy of higher education \cite{gurin2002diversity}.  It has also featured prominently in popular press and legal discourse on higher education.  
Student populations research is widespread in physics education research.
The available student populations influence the way physics education research is conducted.  Ultimately, the effectiveness of educational innovations that are produced by physics education research will depend on how comprehensively student diversity has been investigated.
\subsection{Gender and Sexuality}

The Suffix Tree Clustering algorithm identified gender as one of the top 20 topics in article abstracts, with the labels ``female students'' and ``male and female.''  This was the only cluster with labels about student subpopulations. It contained 26 articles.  The presence of this cluster suggests that gender is the most studied aspect of student populations in \textit{Physical Review Physics Education Research}.  The ease of finding male and female research subjects and extensive concern about the gender gap in student performance are contributors to the presence of this category.  

As shown in Table \ref{tab:gender}, it is common for articles in \textit{Physical Review Physics Education Research} to mention gender terminology.  Many articles use the terminology but do not appear in the gender cluster.  This is because these articles do not feature gender prominently.  For example, articles may mention the gender distribution of research subjects even if it is not part of the experimental design, as a way of characterizing the sample of subjects.
\begin{table}
\caption{\label{tab:gender}Articles referring to gender/sexuality diversity terms.}
\begin{ruledtabular}
\begin{tabular}{lD{.}{.}{-1}r}
\multicolumn{1}{c}{Category}&\multicolumn{1}{c}{Proportion}&\multicolumn{1}{c}{Number}\\
\textbf{gender}&0.38&131\\
\textbf{sexual orientation}&0.03&9\\
\textbf{gender identity}&0.01&2\\
\end{tabular}
\end{ruledtabular}
\end{table}

Unlike gender, sexual orientation and gender identity are not frequently considered in physics education research.  While these widespread \cite{sell1995prevalence,reed2009gender} aspects of diversity are often less conspicuous in the classroom, they are important to student success \cite{meyer2007but}.  Public awareness of sexual orientation has grown greatly since the founding of \textit{Physical Review Physics Education Research} \cite{gallup}.  Recently, gender identity in educational settings has been the subject of intense media, political, and legal scrutiny \cite{dearcol,porter2015assumptions}.  The American Physical Society has released its first report on the inclusion of LGBT professionals in the physics community \cite{lgbtphys}. The connection of students' gender identity and sexual orientation (beyond ``Male or female?'' \cite{kitzinger1999intersexuality,eddy2016beneath,traxler2016enriching}) to physics education should have greater importance to the research community in the future.
\subsection{Race and Ethnicity}
No prominent clusters of articles about students' race or ethnicity were found, but Table \ref{tab:race} shows that these concepts have been considered in a meaningful quantity of physics education research.  Owing to homography, not all aspects of race can be readily identified by terminology frequency analysis.  Like gender, race is often conspicuous in the classroom.  However, race is more conceptually fragmented, which can present challenges to researchers using demographic surveys.  Some groups are neglected in education research in general because of a model minority myth \cite{museus2009deconstructing}.  Indigenous and multiracial student populations are in need of future consideration in physics education research.
\begin{table}
\caption{\label{tab:race}Articles referring to race/ethnicity diversity terms.}
\begin{ruledtabular}
\begin{tabular}{lD{.}{.}{-1}r}
\multicolumn{1}{c}{Category}&\multicolumn{1}{c}{Proportion}&\multicolumn{1}{c}{Number}\\
\textit{hispanic}&0.07&25\\
\textit{african}&0.06&20\\
\textit{native american}&0.02&7\\
\textbf{indigenous}&0&0\\
\textbf{multiracial}&0&1\\
\end{tabular}
\end{ruledtabular}
\end{table}
\subsection{Disability}
\begin{table}
\caption{Articles referring to disability terms.\label{tab:disability}}
\begin{ruledtabular}
\begin{tabular}{lD{.}{.}{-1}r}
\multicolumn{1}{c}{Category}&\multicolumn{1}{c}{Proportion}&\multicolumn{1}{c}{Number}\\
\textit{disab}\{ility\}&0.03&10\\
\textit{anxiet}\{y\}&0.04&12\\
\textit{learning disab}\{ility\}&0.01&2\\
\textit{physical disab}\{ility\}&0&1\\
\textit{dyscalculia}&0&0\\
\textit{dyslexia}&0&0\\
\textit{ableism}&0&0\\
\end{tabular}
\end{ruledtabular}
\end{table}

In the United States, the prevalence of disability is around 22\% \cite{centers2009prevalence}.  Many types of disability have a direct impact on learning physics.  These disabilities are often invisible but widespread.  For instance, it would be unusual to find a class where no students had attention deficit hyperactivity disorder \cite{dupaul2001self}.  Learning disabilities \cite{altarac2007lifetime} and anxieties \cite{betz1978prevalence} are also prevalent.  However, Table \ref{tab:disability} shows disabilities have been considered very little in physics education research, despite their importance to student learning.  Increased awareness among researchers of the prevalence of disability may lead to a greater appreciation of its importance and presence in the student populations being studied.
\subsection{Student Goals}
Alignment between students' goals and their educational experiences is important to ensuring students' motivation and also their satisfaction with schooling \cite{goodlad1984place,kaufman1991influences}.  Students' goals can vary considerably.  For example, some students are seeking preparation for the workplace, while others are looking for personal growth (understanding for its own sake).  Some students strive only for continued enrollment.  These student preferences are an important aspect of student populations.  
Table \ref{tab:goals} suggests that physics education research has considered the professional preparation of students, however not much evidence of research on other types of student goals was found.

\begin{table}
\caption{Articles referring to different types of student goals.\label{tab:goals}}
\begin{ruledtabular}
\begin{tabular}{lD{.}{.}{-1}r}
\multicolumn{1}{c}{Category}&\multicolumn{1}{c}{Proportion}&\multicolumn{1}{c}{Number}\\
\textit{student goal}&0&1\\
\textit{goal congruity}\cite{diekman2010seeking,diekman2013navigating}&0&0\\
\textit{agentic goal}\cite{diekman2010seeking,diekman2013navigating}&0&0\\
\textit{communal goal}\cite{diekman2010seeking,diekman2013navigating}&0&0\\
\textbf{learning orientation}\cite{beaty1997learning}&0.03&9\\
\textit{career}&0.24&82\\
\textit{financ}\{e\}&0.08&26 \\
\textit{business}&0.05&16\\
\textit{industr}\{y\}&0.05&16\\
\textit{personal development}&0&1\\
\textit{personal growth}&0&0\\
\textit{student empower}\{ment\}&0&0\\
\end{tabular}
\end{ruledtabular}
\end{table}

\subsection{Theoretical Viewpoints on the Diversity of Student Populations}

While diversity has played an important role in physics education research, Table \ref{tab:diversitytheory} shows that selected abstract concepts about diversity taken from social sciences have rarely been adopted in physics education research papers.  Some of these concepts are more specific than others, as detailed in the references in Table \ref{tab:diversitytheory}. I will give three examples of how these concepts relate to physics education.

\begin{table}
\caption{\label{tab:diversitytheory}Articles mentioning some theories about diversity and its relationship to education.}
\begin{ruledtabular}
\begin{tabular}{lD{.}{.}{-1}r}
\multicolumn{1}{c}{Category}&\multicolumn{1}{c}{Proportion}&\multicolumn{1}{c}{Number}\\
\textit{affirmative action}&0&0\\
\textbf{critical theor}\{y\} \cite{mezirow1981critical,yanoshak2011educating,brookfield2004power,rosa2016educational}&0&1\\
\textit{internalized oppression} \cite{harper2007peer}&0&0\\
\{in\}\textit{visible minorit}\{y\} \cite{museus2009deconstructing,casas1984profiling}&0.01&2\\
\textit{multicultural}&0.01&2\\
\textit{queer theor}\{y\} \cite{meyer2007but}&0&0\\
\textit{stereotype threat} \cite{madsen2013gender,kreutzer2012preliminary,traxler2015equity}&0.02&7\\
\textbf{universal instruction design} \cite{higbee2008pedagogy}&0&0\\
\end{tabular}
\end{ruledtabular}
\end{table}

\textbf{Universal instruction design} is a proactive framework for educating a diverse group of students \cite{higbee2008pedagogy}.  It seeks to provide for the needs of all students in a unified manner during the design stage of an educational program.  Universal instruction design stands in contrast to the accommodation framework, which reacts to diversity in the classroom by modifying the program of study.  In the physics classroom, an accommodation approach might offer extra knowledge transmission activities in response to the observation of minorities with lower learning gains.  
A universal instruction design approach to learning gaps \cite{kreutzer2012preliminary,madsen2013gender} might replace knowledge transmission pedagogy with peer instruction, which is known to reduce learning gaps \cite{lorenzo2006reducing,pollock2007reducing}.  
Universal instruction design might modify peer instruction further to encourage multisensory  (such as both spoken and written) communication  between students in order to seamlessly include any students  with a sensory impairment.  Benefits of universal instruction design include timeliness and inclusiveness.

\textit{Queer theory} is closely related to gender and feminism.  It is also concerned with how unspoken norms shape learning \cite{meyer2007but}.  
A straightforward application is the hypothesis that norms about gender and sexuality contribute to gaps in learning, but a less direct application of queer theory would be to hypothesize that unspoken social norms contribute to non-Newtonian thinking \cite{halloun1985common,formica2010transforming},  a ``matrix'' \cite{meyer2007but} that physics education researchers are concerned with breaking down.  It seems unlikely that explicit, rather than implicit,  communication is the means of transmission of physics misconceptions.

As an example of the way social interaction might contribute to a non-Newtonian view of inertia, imagine a social situation where two people are moving a wheeled cart and one tells the other to ``keep pushing.'' This interaction might, through an implied social norm, lead to a belief in the concept that objects stop in the absence of force.
Considering these potential links, it is perhaps surprising that queer theory is not widely used in physics education research.  

A \textit{visible minorit}\{y\} is one where membership of the group can be casually identified.  For example, a student who is disabled because they are missing a limb would be visibly disabled, while a student who is color-blind would be invisibly disabled.  It would be unlikely that the student who is color-blind  would be identified as a student who is disabled in a classroom setting unless the student volunteered the information.  The results from Section \ref{sec:pop} suggest that physics education research considers visible groups of students more frequently than invisible groups, and not in proportion to the prevalence of members of the group.

\section{Research Methodology}
The type of methods used in research influences the aspects of student populations that are studied; it is easier to research small minority groups using case studies than to use surveys.  
As in any field, using a range of research methods is important to the success of physics education research.  Multiple methodological approaches are needed because all methodologies have limitations.  For example, some methods are labor intensive, limiting sample size, while others require large sample sizes to be valid.  The sample size required to achieve sufficient statistical power increases when the number of variables in an experiment increases.  This study investigates methodologies used in the journal but does not establish evidence of an interaction between the inclusion of certain methodologies and the inclusion of diversity concepts in a paper.

Survey methodology is frequently mentioned in \textit{Physical Review Physics Education Research} (Table \ref{tab:methods}).  Surveys are likely to appeal to physics education researchers because they can be rapidly repeated, standardized across laboratories, and quantitatively analyzed.  Interviews are also frequently mentioned, while case studies are not uncommon.  Focus groups, which can elucidate the role of social processes \cite{kitzinger1995qualitative} that enhance learning \cite{turpen2009not}, are rarely mentioned.  
\begin{table}
\caption{\label{tab:methods} Articles mentioning research methodology terms.}
\begin{ruledtabular}
\begin{tabular}{lD{.}{.}{-1}r}
\multicolumn{1}{c}{Category}&\multicolumn{1}{c}{Proportion}&\multicolumn{1}{c}{Number}\\
\textit{survey}&0.69&236\\
\textit{interview}&0.62&214\\
\textit{case stud}\{y\}&0.37&125\\
\textit{focus group}&0.03&9\\
\end{tabular}
\end{ruledtabular}
\end{table}

\subsection{Instrumentation}
The Force Concept Inventory (FCI, 53 articles) \cite{hestenes1992force} and Colorado Learning Attitudes about Science Survey (CLASS, 41 articles) \cite{adams2006new} are survey instruments that have associated Suffix Tree Clusters of articles.  These clusters were the second and third highest scoring clusters, scoring slightly higher than ``Physics Education Research.''  This highlights the popularity and influence of survey instruments in the community of authors.  Table \ref{tab:instrument} shows that the Force Concept Inventory is also the most-mentioned instrument in \textit{Physical Review Physics Education Research}. 
\begin{table*}
\caption{\label{tab:instrument}Articles mentioning some common instruments used to measure student learning or attitudes.  Quantitative instrumentation which requires large sample sizes is not suitable for investigations of small minority groups.}
\begin{ruledtabular}
\begin{tabular}{lD{.}{.}{-1}rr}
\multicolumn{1}{c}{Category}&\multicolumn{1}{c}{Proportion}&\multicolumn{1}{c}{Number}&\multicolumn{1}{c}{Reference}\\
\textit{Force Concept}\{ Inventory\}& 0.39&132&\cite{hestenes1992force}\\
\textbf{Motion Conceptual Evaluation}&0.18&62&\cite{thornton1998assessing}\\
\textit{Colorado Learning Attitudes}\{ About Science\}&0.17&57&\cite{adams2006new,adams2004design}\\
\textit{Brief Electricity and Magnetism Assessment}&0.12&40&\cite{ding2006evaluating,chabay1997qualitative}\\
\textit{Maryland Physics Expectation}&0.09&30&\cite{redish1998student}\\
\textit{Conceptual Survey of Electricity and Magnetism}&0.06&20&\cite{maloney2001surveying}\\
\textit{Colorado Upper-Division Electrostatics}&0.04&12&\cite{chasteen2012colorado}\\
\end{tabular}
\end{ruledtabular}
\end{table*}

\begin{figure}
\includegraphics[width=.5\textwidth]{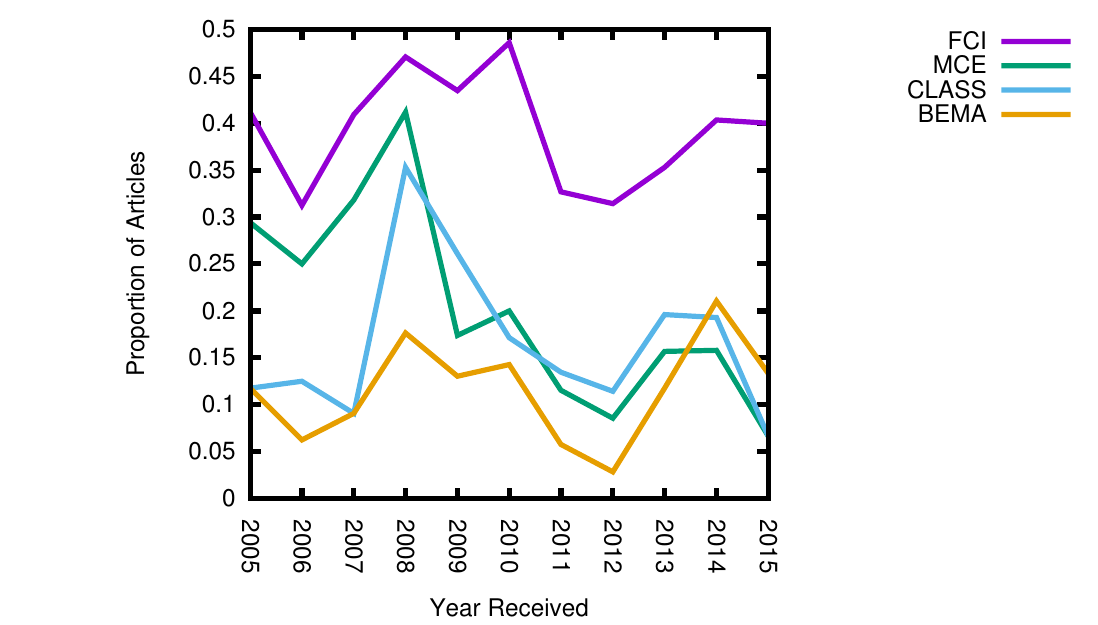}
\caption{\label{fig:instyear}The influential role of the Force Concept Inventory instrument in \textit{Physical Review Physics Education Research} has been maintained throughout the history of the journal.  The proportion of articles mentioning each of four common instruments is displayed by the year the articles were received by the journal.}
\end{figure}

Figure \ref{fig:instyear} shows the evolution of the proportion of articles mentioning selected survey instruments as a function of the year the article was received by the journal.  The instruments predate the founding of the journal.  The number of articles published has been growing rapidly, but on a proportional basis, most instruments have had fairly flat popularity.  The Force Concept Inventory has been consistently popular.  

On the whole, easy-to-use and pre-validated survey instruments are very successful.  However, these instruments are primarily focused on conceptual and attitudinal research, suggesting that there is a need for widely-adopted instruments addressing other important areas, such as quantitative or laboratory skills.  Instruments that focus on a narrow range of reasoning may not address the goals or deficits of all types of students.  The best-known instruments in physics education research are typically analyzed quantitatively.  The widespread adoption of qualitative instruments could enable investigations of smaller or intersectional \cite{traxler2016enriching} minority groups while maintaining a degree of lab-to-lab consistency.

\subsection{Educational Setting}

\subsubsection{Education Level}

\begin{table}
\caption{\label{tab:level}Articles referring to various education levels.}
\begin{ruledtabular}
\begin{tabular}{lD{.}{.}{-1}r}
\multicolumn{1}{c}{Category}&\multicolumn{1}{c}{Proportion}&\multicolumn{1}{c}{Number}\\
\textbf{primary school}&0.11&39\\
\textbf{secondary school}&0.51&174\\
\textbf{introductory physics}&0.91&310\\
\textit{undergrad}\{uate\}&0.61&210\\
\textbf{postgrad}\{uate\}&0.15&51\\
\end{tabular}
\end{ruledtabular}
\end{table}

As Table \ref{tab:level} shows, physics education research considers all education levels. The education level at which physics education research is conducted is a methodological choice which directly impacts the population studied because the education level is strongly correlated with age.  

High school papers formed a Suffix Tree Cluster of 67 abstracts. Upper division papers formed a Suffix Tree Cluster of 36 abstracts.     
The cluster is labeled ``Upper--division,''  and ``Paper is Part.''  
There are nineteen papers in the journal's ``Focused Collection on Upper Division Physics Courses.''  
The label included in abstracts in this collection assisted in the identification of the cluster, but the clustering algorithm successfully identified 14 papers about upper division undergraduate courses that predate the creation of the collection. One paper about upper division secondary students, one paper including ``upper,'' and one paper containing ``paper is part'' were in the cluster.

Tertiary science education requirements for degrees that are not in physics may be influential in shaping physics education research; the \textbf{introductory physics} category is prominent in Table \ref{tab:level}.  Instruments aimed toward the introductory level have popularity far exceeding that of instruments for upper-division education.   Required tertiary science education is an area where physics education research can make a large impact on society.  It is also an area where there is a ready supply of research subjects at many American institutions. 

Research relating to primary school, however, is less common.  In developing nations, some students only have access to primary education\cite{barro2013new}. 
Therefore, physics education research which is pertinent to students at early educational stages can potentially have a broader impact.
\subsubsection{Educational Strategy}

Understanding and improving methods of teaching and learning is the goal of education research. The most powerful way to produce research results which inform practitioners about the impact of their choice of instructional methods on diversity is to conduct experiments which measure the dependence of a learning gain on the  interaction of an instructional method variable and one or more \cite{eddy2016beneath} diversity variables.  To understand the context of diversity studies in physics education research, it is therefore necessary to be aware of the educational strategies which are investigated in the physics education research community.  

The strategies investigated here vary from extremely broad (such as active learning) to highly specific and prescriptive (such as SCALE-UP\cite{beichner2007student}, which recommends a specific diameter for classroom tables).  They also overlap.  The number of articles mentioning a particular teaching strategy should be judged in the context of this variation.  For the more specific instruction methods, references are given in Table \ref{tab:instructionmethod} which give the scope and context of the named educational strategy.

Table \ref{tab:instructionmethod} suggests that \textit{lecture} is the most important instruction method in physics education research.  However, this term is problematic.  To some, it might refer to a knowledge transmission-based instruction.  To others, it might refer to a portion of the course schedule in which a variety of learning modes could take place.  Similarly, \textit{laboratory} could refer to hands-on learning activities performed  by students  or to the place in which research takes place.  Unsurprisingly, both these terms are very common in articles.

\begin{table}
\caption{\label{tab:instructionmethod}Articles referring to some instruction methods and teaching strategies.}
\begin{ruledtabular}
\begin{tabular}{lD{.}{.}{-1}r}
\multicolumn{1}{c}{Category}&\multicolumn{1}{c}{Proportion}&\multicolumn{1}{c}{Number}\\
\textit{lecture}&0.79&270\\
\textit{laboratory}&0.54&185\\
\textit{active learning}&0.38&109\\
\textit{peer instruction}&0.31&107\\
\textit{demonstration}&0.27&93\\
\textit{SCALE-UP} \cite{beichner2007student}&0.08&26\\
\textit{studio physics} \cite{furtak2001installing}&0.07&24\\
\textit{workshop physics} \cite{laws1991calculus}&0.06&22\\
\textit{online course}&0.04&13\\
\textit{inquiry-based learning}&0.01&5\\
\textit{flipped classroom}&0&0\\
\textit{service learning}&0&0\\
\end{tabular}
\end{ruledtabular}
\end{table}

\textit{Active learning} and \textit{peer instruction} are frequently discussed learning modes.  Peer instruction is notable as an example of an instructional technique that has been popularized by physicists, rather than adapted to physics.  On the other hand, the \textit{flipped classroom} has not been mentioned in the journal, though it has grown rapidly in popularity since 2012.  The absence of this term is particularly striking since ``flipped'' instruction can facilitate active learning and peer instruction. Recent work describes different interpretations of ``flipped'' and places the ``flipped classroom'' in the context of common physics education research paradigms\cite{wood2016characterizing}.  \textit{Service learning} \cite{barton2000crafting} is another active learning mode that has gone unmentioned.

\section{Conclusion}
This study is limited in scope to a single journal, which was selected because it attracts articles which contain both education research and relevance to physics.  It also forms a consistently organized corpus.  This is a nonrandom sample which excludes physics education research publications which are targeted to practitioners, education researchers outside physics, and conferences.  The analysis is performed by classifying articles by topic and by searching the text.  The former method detects the core themes of articles but misses the details.  The latter method is sensitive to details but, owing to the inherent properties of natural language, cannot exclude background signals such as proper nouns.  It is also subject to observer bias.

The results show that there is a vigorous and expansive research community backing \textit{Physical Review Physics Education Research}.  This community has shown strong support for student diversity, especially with respect to gender.  It also has a variety of ideas about learning and methods of research.  Like any other field of research, physics education research is not all-encompassing, so I would like to end this article with three points of opinion.

First, invisible student minorities should be studied more.  These are students who are different from the majority, but in a way that is not obvious.  They are numerous, but they are not investigated as much as members of obvious minorities.  Practitioners need to know what works for these students.

Second, students who have disabilities should be studied more.  Some disabilities have a profound impact on learning.  Students with disabilities are not rare, but they are little studied.  

Third, studies of conceptual learning are very popular.  They comprise three of the top twenty Suffix Tree Clusters, and a majority of the widely adopted survey instruments.  Conceptual learning is both difficult and essential; we would not want future leaders to miss it.  However, while conceptual learning is necessary, it is not sufficient for all situations.  In the future, students will be educated using policy recommendations from physics education research.  Physics education research should continue to advance outside the area of conceptual learning to ensure the success of students who will eventually have non-conceptual responsibilities.  That means we need both ``Physics for Future Presidents'' and ``Physics for Future Engineers.''  The students who self-select to pursue such distinct types of physics education may be different enough to need different instructional methods.

Physics education research has benefited from growth in funding and institutional support.  This growth is recent, compared to the history of physics as a whole.  Continued development of diversity in physics education research will benefit from campaigns for more resources.  However, resources are always constrained, so diversity may benefit from an increase in the use of methods like case studies and focus groups, which do not require the recruitment of large numbers of minority students.

\appendix

\section{Search String Lists \label{wl}}

Articles were counted as mentioning the categories labeled in bold face if they mention any of the strings listed in the category.  Text in brackets was not included in the search; it provides example context where the intent of the search might otherwise be unclear.
\begin{enumerate}
\item \textbf{critical theor}: critical theor\{y\}, critical pedagog\{y\}
\item \textbf{gender}: \{fe\}male, gender
\item \textbf{gender identity}: gender identity, transgender, transphob\{ia\}, transsexual
\item \textbf{indigenous}: indigenous, aboriginal
\item \textbf{introductory physics}: first year, introductory physics
\item \textbf{learning orientation}: orientation to learning, learning orientation, orientation toward learning, orientation towards learning
\item \textbf{Motion Conceptual Evaluation}: Motion Conceptual Evaluation, Motion Concept Evaluation
\item \textbf{multiracial}: multiracial, biracial
\item \textbf{postgrad}: doctoral, postgrad\{uate\}
\item \textbf{primary school}: elementary school, primary school
\item \textbf{secondary school}: secondary school, middle school, high school
\item \textbf{sexual orientation}: LGB, queer, lesbian, gay, homosexual, sexual orientation, homophob\{ia\}
\item \textbf{universal instruction design}: universal instruction, universal design
\end{enumerate}

\bibliography{per}
\end{document}